
\input psfig
\magnification=\magstep1
\vsize=47.5pc
\overfullrule=0pt

\def\sn{\smallskip\noindent}
\def\mn{\medskip\noindent}
\def\bn{\bigskip\noindent}
\def\sw{$\cal SW$}
\def\w{\cal W}

\def\o#1{\overline{#1}}

\def\Dt{\Delta}
\def\la{\langle}
\def\ra{\rangle}
\def\lb{\lbrack}
\def\rb{\rbrack}

\def\BZT{{\rm Z{\hbox to 3pt{\hss\rm Z}}}}
\def\BZS{{\hbox{\sevenrm Z{\hbox to 2.3pt{\hss\sevenrm Z}}}}}
\def\BZSS{{\hbox{\fiverm Z{\hbox to 1.8pt{\hss\fiverm Z}}}}}
\def\BZ{{\mathchoice{\BZT}{\BZT}{\BZS}{\BZSS}}}
\def\BQT{\,\hbox{\hbox to -2.8pt{\vrule height 6.5pt width .2pt\hss}\rm Q}}
\def\BQS{\,\hbox{\hbox to -2.1pt{\vrule height 4.5pt width .2pt\hss}$
    \scriptstyle\rm Q$}}
\def\BQSS{\,\hbox{\hbox to -1.8pt{\vrule height 3pt
width .2pt\hss}$\scriptscriptstyle \rm Q$}}

\def\BCT{\,\hbox{\hbox to -3pt{\vrule height 6.5pt width .2pt\hss}\rm C}}
\def\BCS{\,\hbox{\hbox to -2.2pt{\vrule height 4.5pt width .2pt\hss}$
    \scriptstyle\rm C$}}
\def\BCSS{\,\hbox{\hbox to -2pt{\vrule height 3.3pt
width .2pt\hss}$\scriptscriptstyle \rm C$}}

\def\BRT{{\rm I{\hbox to 5.5pt{\hss\rm R}}}}
\def\BRS{{\hbox{\sevenrm I{\hbox to 4.3pt{\hss\sevenrm R}}}}}
\def\BRSS{{\hbox{\fiverm I{\hbox to 3.35pt{\hss\fiverm R}}}}}
\def\BR{{\mathchoice{\BRT}{\BRT}{\BRS}{\BRSS}}}
\def\BNT{{\rm I{\hbox to 5.5pt{\hss\rm N}}}}
\def\BNS{{\hbox{\sevenrm I{\hbox to 4.3pt{\hss\sevenrm N}}}}}
\def\BNSS{{\hbox{\fiverm I{\hbox to 3.35pt{\hss\fiverm N}}}}}
\def\BN{{\mathchoice{\BNT}{\BNT}{\BNS}{\BNSS}}}
\def\q#1{\lbrack#1\rbrack}
\def\section#1{\leftline{\bf #1}\vskip-7pt\line{\hrulefill}}
\def\bibitem#1{\parindent=8mm\item{\hbox to 6 mm{$\q{#1}$\hfill}}}
\def\bou{1}
\def\ina{2}
\def\lup{3}
\def\rom{4}
\def\nem{5}
\def\ito{6}
\def\blm{7}
\def\ban{8}
\def\gep{9}
\def\kaz{10}
\def\ber{11}
\def\var{12}
\def\ehe{13}
\def\rav{14}
\def\qiu{15}
\def\odz{16}
\def\egu{17}
\def\wes{18}
\def\mus{19}
\def\hon{20}
\def\gin{21}
\def\kir{22}
\def\flo{23}
\def\zam{24}
%
%
\font\large=cmbx10 scaled \magstep3

\font\bigf=cmr10 scaled \magstep2
\pageno=0
\def\folio{
\ifnum\pageno<1 \footline{\hfil} \else\number\pageno \fi}
\phantom{not-so-FUNNY}
\rightline{ BONN--TH--94--08\break}
\rightline{ hep-th/9407068\break}
\rightline{ July 1994\break}
\vskip 2.0truecm
\centerline{\large A Note on Representations}
\vskip 0.3truecm
\centerline{\large of N=2 SW-algebras}
\vskip 1.0truecm
\centerline{\bigf R.\ Blumenhagen, R.\ H\"ubel}
\bigskip\medskip
\centerline{\it Physikalisches Institut der Universit\"at Bonn}
\centerline{\it Nu{\ss}allee 12, 53115 Bonn, Germany}
\medskip
\vskip 1.1truecm
\centerline{\bf Abstract}
\vskip 0.2truecm
\noindent
We investigate the representation theory of some recently
constructed $N=2$ super $\w$-algebras with two generators.
Except for the central charges in the unitary minimal series
of the $N=2$ super Virasoro algebra we find no new rational
models. However, from our results it is possible to arrange
all known $N=2$ super $\w$-algebras with two generators and
vanishing self-coupling constant into four classes.
For the algebras existing for $c\ge 3$ which can be understood
by the spectral flow of the $N=2$ super Virasoro algebra we
find that the representations have quantized $U(1)$ charge.
\vfill
\eject
\section{1.\ Introduction}
\sn
In the last few years the explicit construction of $\w$-algebras
$\q{\bou}$ has also reached the fast developing field of $N=2$
supersymmetric conformal field theories (SCFTs) \q{\ina-\blm}.
{}From a two dimensional quantum-gravity point of view $N=2$ SCFTs
are very important since on the one hand they describe the
string world sheet CFT of $N=1$ space-time supersymmetric heterotic
string compactifications \q{\ban-\kaz} and on the
other hand the ghost sector of $\w$-gravity theories contains an
$N=2$ super $\w$-algebra \q{\ber}. In ref.\ $\q{\blm}$ non-linear
$N=2$ super $\w$-algebras, namely extensions of
the $N=2$ super Virasoro algebra ($N=2$ SVIR) by a pair of
super-primary fields of opposite $U(1)-$charge, have been treated in
a systematic way. The calculation of such objects is very involved but
using a manifestly covariant approach one can simplify the problem
tremendously. The extension of $N=2$ SVIR by superprimaries of
superconformal dimension $\Dt_1,\ldots,\Dt_n$ is denoted by
${\cal SW}(1,\Dt_1,\ldots,\Dt_n)$.
\sn
After presenting a further example, the ${\cal SW}(1,3)$ algebra
with zero $U(1)-$charge and non-vanishing self-coupling constant,
we investigate the representation theory of the former algebras
focusing on the question whether these algebras admit rational models.
To this end we evaluate Jacobi identities on the lowest weight
states $|h,q\ra$ yielding restrictions on the allowed
lowest weights $(h,q)$ \q{\var,\ehe}.
The results obtained by this method allow one to arrange all
known $N=2$ super $\w$-algebras with two generators and
vanishing self-coupling constant into essentially four series.
Only one series contains rational theories which can be explained by the
classification of modular invariants of $N=2$ SVIR \q{\rav,\qiu}.
In the case of $N=2$ super $\w$-algebras with two generators there are
many examples of algebras existing for $c\ge 3$.
All these $c\ge 3$ theories have common features which can be
explained by the existence of the spectral flow of $N=2$ SVIR.
Using this fact, we can describe these $N=2$ super $\w$-algebras by a simple
formula containing the well known algebras ${\cal SW}(1,{d\over 2})$
for $Q=d$, $c=3d$ ($d\in\BN$) $\q{\odz}$ as a subset.
The latter algebras occur in the compactification of the heterotic
string to $(10-2d)$ space-time dimensions~\q{\egu,\odz}.
\sn
\section{2.\ Construction of $N=2$ super $\w$-algebras: \sw(1,3)}
\sn
In this section we use results of \q{\wes,\mus} about
holomorphic $N=2$ superconformal field theories, and in particular
we will use the notation and formulae of section two of
ref.\ \q{\blm}. We apply the formalism of \q{\blm} to construct
the supplementary example \sw(1,3) with zero $U(1)-$charge and
non-vanishing self-coupling constant. Firstly, we have to write down all
super quasiprimary fields which can occur in the OPE
$\Phi_{3}^{0}(Z_1)\>\Phi_{3}^{0}(Z_2)$.
We present only the fields of dimension 5 and ${11}\over 2$.
The fields with lower dimension have been presented in~$\q{\blm}$.
\sn
\centerline{\vbox{\hbox{\vbox{\offinterlineskip
\def\tablespace{height2pt&\omit&&\omit&&\omit&\cr}
\def\tablerule{\tablespace\noalign{\hrule}\tablespace}
\hrule\halign{&\vrule#&\strut\hskip0.2cm\hfil#\hfill\hskip0.15cm\cr
\tablespace
& $\Dt$ && $Q$ && quasi-primary fields &\cr
\tablerule\tablerule
& 5 && 0 && ${\cal N}_s({\cal N}_s({\cal N}_s({\cal N}_s({\cal L}{\cal L})
{\cal L}){\cal L}){\cal L}),\
{\cal N}_s({\cal N}_s({\cal N}_s({\cal L}[\o{D},D]{\cal L})
{\cal L}){\cal L}),\
{\cal N}_s({\cal N}_s({\cal L}\partial^2{\cal L}){\cal L})$ &\cr
\tablespace
& \omit && \omit &&  ${\cal N}_s({\cal N}_s({\cal L}\o{D}
\partial{\cal L})D{\cal L}),
\ {\cal N}_s({\cal N}_s({\cal L}D\partial{\cal L})\o{D}{\cal L}),\
{\cal N}_s({\cal L}[\o{D},D]\partial^2{\cal L})$ &\cr
\tablespace
& \omit && \omit && ${\cal N}_s({\cal N}_s({\Phi}{\cal L}){\cal L}),\
           {\cal N}_s({\Phi}\partial{\cal L}),\
           {\cal N}_s({\Phi}[\o{D},D]{\cal L})$ &\cr
\tablespace
& $11\over2$ && -1 &&
${\cal N}_s({\cal N}_s({\cal N}_s({\cal L}{D}
\partial{\cal L}){\cal L}){\cal L}),
\ {\cal N}_s({\cal N}_s({\cal N}_s({\cal N}_s({\cal L}{\cal L})
{\cal L}){\cal L})D{\cal L}),\
{\cal N}_s({\cal N}_s({\cal L}\partial^2{\cal L})D{\cal L})$ &\cr
\tablespace
& \omit && \omit &&
${\cal N}_s({\cal N}_s({\cal N}_s({\cal L}[\o{D},D]{\cal L})
{\cal L})D{\cal L}),\ {\cal N}_s({\Phi}\partial{D}{\cal L}),
\ {\cal N}_s({\cal N}_s({\Phi}{D}{\cal L}){\cal L})$ &\cr
\tablespace
& $11\over2$ && 1 &&
 ${\cal N}_s({\cal N}_s({\cal N}_s({\cal L}\o{D}
\partial{\cal L}){\cal L}){\cal L}),
\ {\cal N}_s({\cal N}_s({\cal N}_s({\cal N}_s({\cal L}{\cal L})
{\cal L}){\cal L})\o{D}{\cal L}),\
{\cal N}_s({\cal N}_s({\cal L}\partial^2{\cal L})\o{D}{\cal L})$ &\cr
\tablespace
& \omit &&\omit && ${\cal N}_s({\cal N}_s({\cal N}_s({\cal L}
                   [\o{D},D]{\cal L}){\cal L})\o{D}{\cal L}),\
 {\cal N}_s({\Phi}\partial\o{D}{\cal L}),\
 {\cal N}_s({\cal N}_s({\Phi}\o{D}{\cal L}){\cal L})$ &\cr
\tablespace}\hrule}}
\hbox{\hskip 0.5cm Table 1: Quasi-primary fields of dimension 5
and ${11 \over 2}$ in ${\cal SW}(1,3)$} }}
\bn
For completeness we give also the Kac-determinants in the vacuum sector:
\sn
\centerline{\vbox{
\hbox{\vbox{\offinterlineskip
\def\tablespace{height2pt&\omit&&\omit&\cr}
\def\tablerule{\tablespace\noalign{\hrule}\tablespace}
\hrule\halign{&\vrule#&\strut\hskip0.2cm\hfil#\hfill\hskip0.2cm\cr
\tablespace
& $\Delta$ && $det(D_{\Delta})\sim$  &\cr
\tablerule\tablerule
& 5 && $(c-2)(c-1)^5 c^9 (c+1)(c+6)^3(c+12)
            (2c-3)^3 (5c-9) (7c+18) (c^2+26c-75)$ &\cr
\tablespace
&$11\over2$&&$(c-1)^6 c^{14}(c+1)^2 (c+6)^4 (2c-3)^4 (5c-9)^2 (7c-18)^2$&\cr
\tablespace}\hrule}}
\hbox{\hskip 0.5cm Table 2: Kac-determinants}}}
\mn
The next step is to determine the structure constants
$C_{ij}^k$, $\alpha_{ijk}$ for all normal
ordered products which are rational functions in $c$ and one
self-coupling constant $C_{33}^3\,\alpha$. We skip their presentation
and continue with the result that the Jacobi identities
yield the following expression for the coupling constant $C_{33}^3\,\alpha$:
$$\bigl(C_{33}^3\,\alpha\bigr)^2
={49c^3(7c-18)^2(c^2+26c-75)(3c^4+62c^3-129c^2-9c+54)\over
   3(3-2c)(c-2)(c-1)(c+1)(c+6)(c+12)(5c-9)(3c^2-37c+60)}.
\eqno{\rm (1)}$$
All Jacobi identities are satisfied only if the central charge takes the
following rational numbers:
$$ c\in\left\lbrace{18\over7},{10\over3},-{9\over5}\right\rbrace.
\eqno{\rm (2)}$$
The corresponding values of the self-coupling constant are
$$ \bigl(C_{33}^3\,\alpha\bigr)^2\in\left\lbrace 0,{268960000\over88803},
          -{106815267\over88000}\right\rbrace.
\eqno{\rm (3)}$$
\mn
We conclude this section with a very brief review of the results of the
construction of $N=2$ super $\w$-algebras with two generators
with {\it vanishing self-coupling constant} \q{\blm}.
Using $\rm MATHEMATICA^{TM}$ and a special C-program \q{\hon}
we constructed ${\cal SW}(1,\Delta)$ algebras for
$\Delta\in\lbrace {3\over2},2,{5\over2},3\rbrace$.
In the following table we present the central charges $c$ and
$U(1)-$charges $Q$ for which these $\w$-algebras exist.
\mn
\centerline{\vbox{
\hbox{\vbox{\offinterlineskip
\def\tablespace{height2pt&\omit&&\omit&&\omit&&\omit
                          &&\omit&&\omit&&\omit&&\omit&\cr}
\def\tablerule{\tablespace\noalign{\hrule}\tablespace}
\hrule\halign{&\vrule#&\strut\hskip0.6cm\hfil#\hfil\hskip0.6cm\cr
\tablespace
&  $\Delta$ && $Q$ && $c$ && \omit && \omit&& $\Delta$ && $Q$ && $c$ &\cr
\tablerule\tablerule
& \omit && $0$ && $9\over4$, $3\over5$, $-{3\over2}$ && \omit&& \omit
&&  \omit  && $0$ && $5\over2$, $9\over7$, $-{9\over2}$  &\cr
\tablespace
& $3\over2$ && $\pm 3$ && $9$ &&\omit&&\omit&&$5\over2$&&$\pm 5$&& $15$&\cr
\tablespace
& \omit && $\pm 1$ && $3$ &&\omit&&\omit&&\omit&& $\pm 1$ && $3$  &\cr
\tablespace
& \omit && \omit&&\omit&&\omit&&\omit&& \omit&&$\pm 2$ && $9\over2$ &\cr
\tablerule
& \omit && 0 && $12\over5$, $-3$&&\omit&&\omit&&\omit&&$0$&&$18\over7$&\cr
\tablespace
&  2 && $\pm 4$  &&  $12$, $-9$, $-21$
&&\omit&& \omit &&  $3$ && $\pm 6$ && $18$, $-15$, $-33$ &\cr
\tablespace
& \omit && $\pm {3\over2}$  &&  $15\over4$
&& \omit&& \omit && \omit && $\pm {4\over3}$ && $10\over3$ &\cr
\tablespace
& \omit  && \omit && \omit  && \omit
&& \omit && \omit&& $\pm {5\over2}$ && $21\over4$ &\cr
\tablespace}\hrule}}
\hbox{\hskip 0.1cm Table 3: Central charges of
${\cal SW}(1,\Delta)$ algebras with vanishing self-coupling constant}}}
\vfill
\eject
\section{3.\ Representations of $N=2$ super $\w$-algebras}
\mn
In this section we study the possible lowest weight representations
in the Neveu-Schwarz (NS) sector of the algebras presented in table 3.
To this end one evaluates Jacobi-identities on the lowest weight
states yielding necessary conditions for the quantum numbers of the
allowed representations. According to the results obtained this way, we
propose an arrangement of $N=2$ super $\w$-algebras with two generators
and vanishing self-coupling constant into the following four classes:
\bn
{\bf (1)} $c={6\Dt\over 2\Dt+1},Q=0,\quad 2\Delta\in\BN$
\sn
These values of $c$ lie in the unitary minimal
series $c(k)={{3k}\over {k+2}},\ k\in \BN$ of $N=2$ SVIR with $k=4\Dt$.
The lowest weights in the NS sector are given by (see e.g.\ $\q{\qiu}$):
$$ h_{l,m}={{l(l+2)-m^2}\over{4(k+2)}},\quad q_m={m\over{k+2}},\quad
l=0,\dots,k\quad {\rm and\ } m=-l,-l+2,\dots,l.
\eqno{\rm (4)}$$
The primary field corresponding to the lowest weight $(h_{k,0},0)$
is a $\BZ_2$ simple current of conformal dimension $\Dt$ for $k=4\Dt$.
The classification of modular invariant partition functions
$\q{\qiu}$ yields off-diagonal invariants for
$k\equiv 0,2 \ ({\rm mod}\ 4)$ which can be viewed as diagonal invariants
of the extended model with symmetry algebra ${\cal SW}(1,\Dt)$.
The explicit calculation of the lowest weights of the ${\cal SW}(1,\Dt)$
at $c(4\Dt)$ yields exactly those lowest weights which one expects from the
off-diagonal invariant. We stress that these models are the only rational
ones among the central charges given in table 3 above.
The NS part of the non-diagonal modular invariant partition function for
$k\equiv 2\ ({\rm mod}\ 4)$ is given by $\q{\qiu}$:
$$\eqalign{Z_{NS}=&
 \sum_{{l=0,2,\dots,{k\over 2}-1}\atop{m=-l,-l+2,\dots,l}}
\hskip -0.5cm {\textstyle{1\over 2}}\vert \chi^l_m+\chi^{k-l}_m \vert^2  +
 \sum_{{l={k\over 2}+1,\dots,k-2,k}\atop{|m|< k-l}}
\hskip -0.5cm {\textstyle{1\over 2}}\vert \chi^l_m+\chi^{k-l}_m \vert^2 +\cr
& \sum_{{l={k\over 2}+1,\dots,k-2,k}\atop{m<-(k-l)}}
\hskip -0.5cm {\textstyle{1\over 2}}\vert \chi^l_m+\chi^l_{m+k+2} \vert^2 +
 \sum_{{l={k\over 2}+1,\dots,k-2,k}\atop{m> k-l}}
\hskip -0.5cm {\textstyle{1\over 2}}\vert\chi^l_m+\chi^l_{m-(k+2)}\vert^2.\cr
}\eqno{\rm (5)}$$
The $\chi^l_m$ are the characters of the
irreducible lowest weight representations of $N=2$ SVIR with
lowest weight $(h_{l,m},q_m)$. For $k\equiv 0\ ({\rm mod}\ 4)$
it reads $\q{\qiu}$:
$$\eqalign{Z_{NS}=&
 \sum_{{l=0,2,\dots,{k\over 2}-2}\atop{m=-l,-l+2,\dots,l}}
\hskip -0.5cm {\textstyle{1\over 2}}\vert \chi^l_m+\chi^{k-l}_m \vert^2 +
 \sum_{m=-{k\over 2},-{k\over 2}+2\dots,{k\over 2}}
\hskip -0.5cm 2\vert \chi^{k\over 2}_{m} \vert^2  +
 \sum_{{l={k\over 2}+2,\dots,k-2,k}\atop{|m|< k-l}}
\hskip -0.5cm {\textstyle{1\over 2}}\vert \chi^l_m+\chi^{k-l}_m\vert^2 +\cr
& \sum_{{l={k\over 2}+2,\dots,k-2,k}\atop{m<-(k-l)}}
\hskip -0.5cm {\textstyle{1\over 2}}\vert \chi^l_m+\chi^l_{m+k+2}\vert^2 +
  \sum_{{l={k\over 2}+2,\dots,k-2,k}\atop{m> k-l}}
\hskip -0.5cm {\textstyle{1\over 2}}\vert\chi^l_m+\chi^l_{m-(k+2)}\vert^2.\cr
}\eqno{\rm (6)}$$
\sn
{}From these expressions one can directly read off the lowest weights
of ${\cal SW}(1,\Dt)$ in the NS-sector.
The following picture shows the allowed lowest weights $(h,q)$ for
the first member of the series $(k=6)$. A cross represents a representation
module with non-degenerate lowest weight, i.e.\ the zero modes of
bosonic components act trivial in the NS sector. The rectangles denote
a representation module with doubly degenerate lowest weight, i.e.\ one
has $\vert h,-{1\over 2}\ra \sim \o{\psi}_0 \vert h, {1\over 2}\ra$
\footnote{${}^1)$}{The components $\phi,\psi,\o\psi,\chi$ of a super
field $\Phi$ are defined according
to $\Phi(Z)=\phi(z)+{1\over{\sqrt{2}}}\left(\theta\o\psi(z)-\o\theta\psi(z)
\right) + \theta\o\theta\chi(z)$.}.
\mn
\vbox{
\line{\hfill\psfig{figure=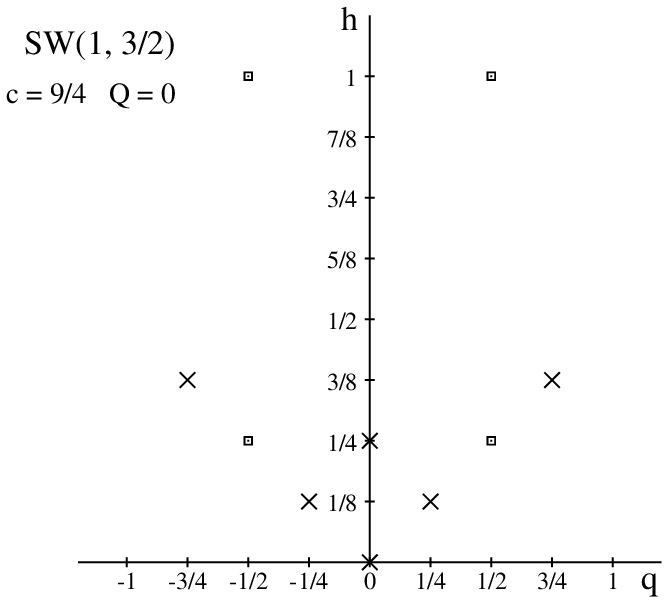}\hfill}
\vskip -3.5cm
\line{\hskip3.5cm $\scriptstyle{\vert h,-{1\over2}\ra
\sim \o{\psi}_0\vert h,{1\over2}\ra}$}}
\vskip 3cm
\bn
The Ramond sector is determined by the spectral flow
$$\eqalign{
L'_n &=L_n+\eta J_n+{c\over 6}\eta^2\delta_{n,0}\ ,\quad
             J'_n=J_n+{c\over 3}\eta\delta_{n,0} \cr
             \varphi'_r&=\varphi_{r+\eta Q },\quad\quad
  \varphi\in\lbrace G,\o{G},{\rm components\ of\ \Phi}\rbrace \cr}
\eqno{\rm (7)}$$
\sn
for $\eta=\pm {1\over 2}$. Note that in the Ramond sector the bosonic fields
$\psi$ and $\bar\psi$ of dimension 2 carry half-integer modes.
\bn\mn
{\bf (2a)} $c=3\left(1-{2\over \Dt+1}\right),Q=0\quad
            {\rm for}\quad \Dt\in \BN+{1\over 2}$
\mn
In particular, the values $c={3\over5}$ ($\Dt={3\over 2}$)
and $c={9\over7}$ ($\Dt={5\over 2}$) are contained in this series.
These models have {\it infinitely} many lowest weight
representations with infinitely degenerate lowest weight.
One obtains for the minimal lowest weight
$h_{\rm min}={c-3\over24}$, i.e.\ one has
$c_{\rm eff}=c-24 h_{\rm min}=3$.
However, there exist finitely many representations
with finitely degenerate lowest weight.
Note that the $N=0$ ($c_{\rm eff}=1$) and $N=1$
($c_{\rm eff}={3\over 2}$) analogues are rational
theories (see e.g.\ \q{\gin-\flo}).
\bn
For $\Dt\in \BN$ there are two other
series with $c_{\rm eff}=3$.
\bn
{\bf (2b)} $c=3\left(1-2\Dt\right),Q=2\Dt$\ \  and\ \
 $c=3\left(1-4\Dt\right),Q=2\Dt$\ \ \ \ for $\Dt\in \BN$
\mn
The space of representations looks similar to the case (2a).
\vfill\eject
\sn
As examples we present the $c={3\over5}$ ($\Dt={3\over 2}$)
and $c=-9$ ($\Dt=2$) models:
\sn
\vbox{
\line{\quad\psfig{figure=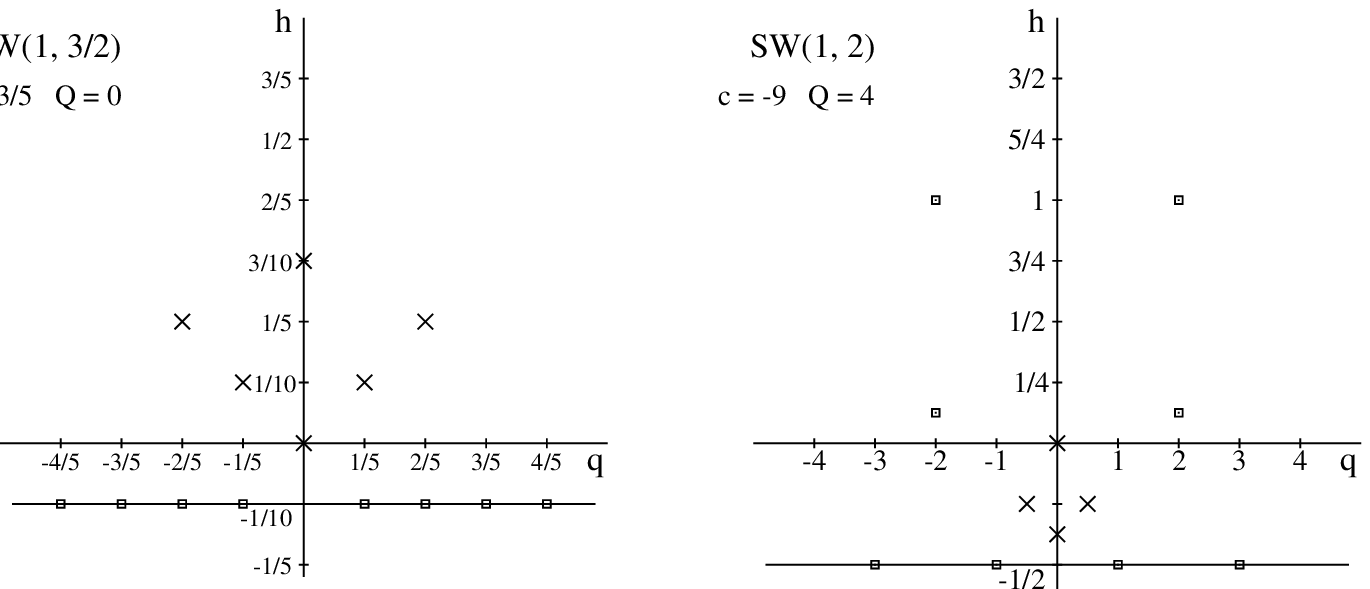}\hfill}
\vskip -4.5cm
\line{\hskip 10cm\hbox{$\scriptstyle
\vert h,-2\ra\sim\varphi_0^-\vert h,2\ra,\ h>0$}\hfil}
}
\vskip 4cm
\bn\mn
Explicitly we obtained the following representations
for ${\cal SW}(1,{3\over 2})$ and $c={3\over 5}$:
\mn
\centerline{\vbox{\hbox{\vbox{\offinterlineskip
\def\tablespace{height2pt&\omit&&\omit&\cr}
\def\tablerule{\tablespace\noalign{\hrule}\tablespace}
\hrule\halign{&\vrule#&\strut\hskip0.2cm\hfill#\hfill\hskip0.2cm\cr
\tablespace
& $(h,q)$ && \omit  &\cr
\tablerule\tablerule
& $(0,0),({3\over{10}},0),
({1\over 5},\pm{2\over 5}),({1\over{10}},\pm{1\over 5})$ &&
$\la h,q\vert \psi_0 \o{\psi}_0 \vert h,q\ra =
\la h,q\vert \o{\psi}_0 {\psi}_0 \vert h,q\ra = 0$ &\cr
\tablerule
& $(-{1\over{10}},q)\quad {\rm for\ }q\in\BR$ &&
$\la h,q\vert \psi_0 \o{\psi}_0 \vert h,q\ra\sim
(1-5q)(2-5q)(3-5q)(4-5q) $ &\cr
&\omit && $\la h,q\vert \o{\psi}_0 {\psi}_0 \vert h,q\ra \sim
(1+5q)(2+5q)(3+5q)(4+5q) $ &\cr
\tablespace}\hrule}}
\hbox{\hskip 0.5cm Table 4:
${\cal SW}(1,{3\over 2}) {\rm \ at\ } c= {3\over 5},Q=0$}}}
\mn
The relations for $h=-{1\over {10}}$ imply that for
$(\pm q)\in \{ {1\over 5}, {2\over 5}, {3\over 5}, {4\over 5} \}$
the degeneracy of the
lowest weight is partially removed but is nevertheless infinite.
That is to say that e.g.\ the states $\vert -{1\over {10}}, -{4\over 5}\ra$
and  $\vert -{1\over{10}}, {1\over 5}\ra$ are not in the same representation
module as it is the case for $\vert -{1\over{10}}, q \ra$
and $\vert -{1\over{10}}, q\pm 1 \ra$ if $q$ is generic.
\bn
For ${\cal SW}(1,2)$ and $c=-9$ we obtained the following results:
\mn
\centerline{\vbox{\hbox{\vbox{\offinterlineskip
\def\tablespace{height2pt&\omit&&\omit&\cr}
\def\tablerule{\tablespace\noalign{\hrule}\tablespace}
\hrule\halign{&\vrule#&\strut\hskip0.2cm\hfill#\hfill\hskip0.2cm\cr
\tablespace
& $(h,q)$ && \omit  &\cr
\tablerule\tablerule
& $(0,0),(-{3\over 8},0),
(-{1\over 4},\pm{1\over 2})$ &&
$\la h,q\vert \phi_0^+ \phi_0^- \vert h,q\ra=
\la h,q\vert \phi_0^- \phi_0^+ \vert h,q\ra = 0$ &\cr
\tablerule
& $({1\over 8},\pm 2)$ &&
$\la h,q\vert \phi_0^\pm \phi_0^\mp \vert h,q\ra \not=0 ,\
\la h,q\vert \phi_0^\mp \phi_0^\pm \vert h,q\ra = 0$ &\cr
\tablerule
& $(1,\pm 2)$ &&
$\la h,q\vert \phi_0^\pm \phi_0^\mp \vert h,q\ra \not= 0,\
\la h,q\vert \phi_0^\mp \phi_0^\pm \vert h,q\ra = 0$ &\cr
\tablerule
& $(-{1\over 2},q)\quad {\rm for\ } q\in\BR$ &&
$\la h,q\vert \phi_0^+ \phi_0^- \vert h,q\ra \sim
(q-3)^2(q-1)^2$ &\cr
& \omit &&
$\la h,q\vert \phi_0^- \phi_0^+ \vert h,q\ra \sim
(q+3)^2(q+1)^2$ &\cr
\tablespace}\hrule}}
\hbox{\hskip 0.5cm Table 5: ${\cal SW}(1,2) {\rm \ at\ } c=-9,Q=4$}}}
\mn
The implication of the relations for $h=-{1\over 2}$
is similar to the previous case. Note that the
representations with positive $h$ are doubly degenerate.
\vfill
\eject
\sn
{\bf (3)} $c=3\left(1-\Dt\right),Q=0,\quad 2\Delta\in\BN$
\bn
These models are pathological because their spectrum
is not bounded from below.
\bn
\vbox{
\line{\hfill\psfig{figure=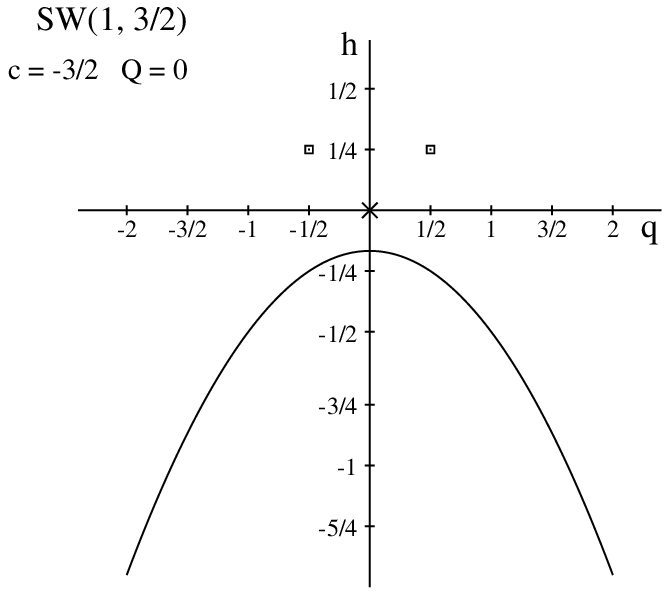}\hfill}
\vskip -5cm
\line{\hskip 8cm\hbox{$\scriptstyle
\vert {1\over 4},-{1\over 2}\ra\sim
{\o\psi}_0\vert {1\over 4},+{1\over 2}\ra$}\hfil}
}
\vskip 4.5cm
\bn
To be explicit, from our calculations we could not exclude the
following representations:
\bn
\centerline{\vbox{\hbox{\vbox{\offinterlineskip
\def\tablespace{height2pt&\omit&&\omit&\cr}
\def\tablerule{\tablespace\noalign{\hrule}\tablespace}
\hrule\halign{&\vrule#&\strut\hskip0.2cm\hfill#\hfill\hskip0.2cm\cr
\tablespace
& $(h,q)$ && \omit  &\cr
\tablerule\tablerule
& $(0,0) ,\  (-{1\over 3}q^2-{1\over 6},q)\quad {\rm for\ }q\in\BR$ &&
$\la h,q\vert \psi_0 \o{\psi}_0 \vert h,q\ra =
\la h,q\vert \o{\psi}_0 {\psi}_0 \vert h,q\ra = 0$ &\cr
\tablerule
& $({1\over 4},+{1\over 2})$ &&
$\la h,q\vert \o{\psi}_0 {\psi}_0 \vert h,q\ra = 0,\
 \la h,q\vert \psi_0 \o{\psi}_0 \vert h,q\ra \not= 0$ &\cr
\tablerule
&  $({1\over 4},-{1\over 2})$ &&
$\la h,q\vert \o{\psi}_0 {\psi}_0 \vert h,q\ra \not= 0,\
 \la h,q\vert \psi_0 \o{\psi}_0 \vert h,q\ra = 0$ &\cr
\tablespace}\hrule}}
\hbox{\hskip 0.5cm Table 6: ${\cal SW}(1,{3\over 2})\ {\rm at\ }
       c=-{3\over 2},Q=0$  }}}
\bn\bn
{\bf (4)} $c\ge3, Q {\rm \ rational}$
\bn
For these algebras there exists no known analogue in the
theory of $N=0$ and $N=1$ $\w$-algebras with two generators
(see e.g.\ \q{\bou} and references therein).
All representations of these algebras have a similar structure.
The theories are not rational and have the property that infinitely
many representations with fixed $U(1)$-charge $q$ exist, where
$q$ takes finitely many rational values. The representations with fixed
$h$ have a finitely degenerate lowest weight.
Furthermore, in all cases the field ${\cal N}_s(\Phi{\cal L})$
vanishes identically.
\mn
For the subset $c=3d$ ($d\in\BN$) it has been shown explicitly in \q{\odz}
that the extension of the $N=2$ SVIR by the local chiral
primaries $(\Delta,Q)=(c/6,\pm c/3)$ exists.
In $\q{\odz}$ the null field ${\cal N}_s(\Phi{\cal L})$ has been
used to find the unitary representations for $c=3d$.
Using the methods described above we were able to reproduce
these results without exploiting the null field:
\vfill\eject
\sn
\vbox{
\line{\hfill\psfig{figure=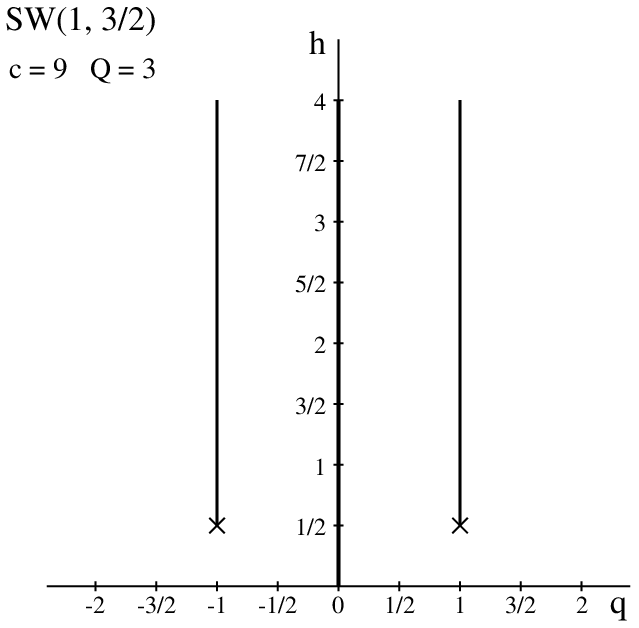}\quad
\quad \quad\psfig{figure=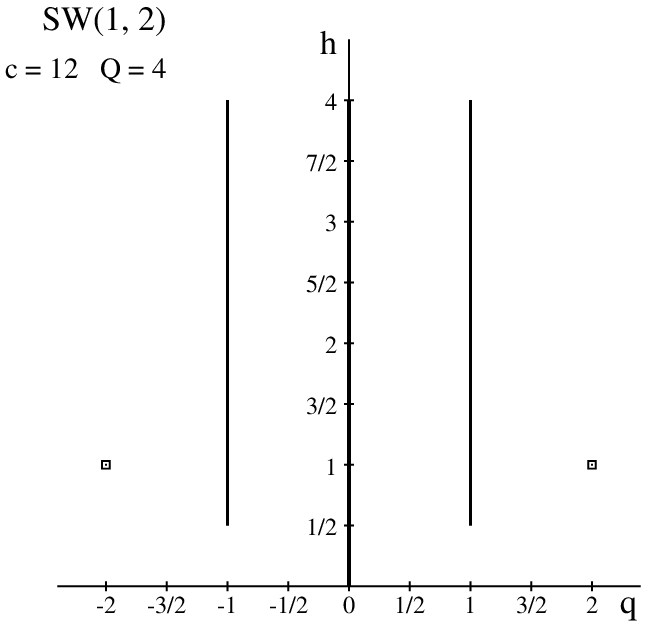}\hfill}
\vskip -4.5cm
\line{\hbox{$\scriptstyle
\vert h,-1\ra\sim{\psi}_0^-\vert h,1\ra,\ h>{1\over 2}$}
\hskip 8.5cm
\hbox{$\scriptstyle \vert 1,-2\ra\sim{\phi}_0^-\vert 1,2\ra$}
\hfil}}
\vskip 4.5cm
\sn
The following tables show the concrete results in more detail.
Note that we give only the unitary irreducible representations
implying the condition $h\ge {1\over 2}\vert q\vert$.
\sn
\centerline{\vbox{\hbox{\vbox{\offinterlineskip
\def\tablespace{height2pt&\omit&&\omit&\cr}
\def\tablerule{\tablespace\noalign{\hrule}\tablespace}
\hrule\halign{&\vrule#&\strut\hskip0.2cm\hfill#\hfill\hskip0.2cm\cr
\tablespace
& $(h,q)$ && \omit &\cr
\tablerule\tablerule
&$({1\over 2},+1),\ (h,-1),(h,0)\quad{\rm for\ }{{|q|}\over 2}\le h<\infty$&&
$\la h,q\vert \o{\psi}_0^+ {\psi}_0^- \vert h,q\ra = 0$ &\cr
\tablerule
&$({1\over 2},-1),\ (h,0),(h,+1)\quad{\rm for\ }{{|q|}\over 2}\le h<\infty$&&
$\la h,q\vert {\psi}_0^- \o{\psi}_0^+ \vert h,q\ra = 0 $ &\cr
\tablespace}\hrule}}
\hbox{\hskip 0.5cm Table 7: ${\cal SW}(1,{3\over 2})$
     {\rm at }$c=9,Q=3$}}}
\sn
\centerline{\vbox{\hbox{\vbox{\offinterlineskip
\def\tablespace{height2pt&\omit&&\omit&\cr}
\def\tablerule{\tablespace\noalign{\hrule}\tablespace}
\hrule\halign{&\vrule#&\strut\hskip0.2cm\hfill#\hfill\hskip0.2cm\cr
\tablespace
& $(h,q)$ && \omit &\cr
\tablerule\tablerule
& $(h,0),(h,\pm 1)\quad {\rm for\ } {{|q|}\over 2}\le h<\infty$ &&
$\la h,q\vert \phi_0^+ \phi_0^- \vert h,q\ra =
            \la h,q\vert \phi_0^- \phi_0^+ \vert h,q\ra = 0$  &\cr
\tablerule
& $(1,-2)$ && $\la h,q\vert \phi_0^+ \phi_0^- \vert h,q\ra = 0,\
\la h,q\vert \phi_0^- \phi_0^+ \vert h,q\ra \not= 0$  &\cr
\tablerule
&  $(1,+2)$ && $\la h,q\vert \phi_0^+ \phi_0^- \vert h,q\ra \not= 0,\
\la h,q\vert \phi_0^- \phi_0^+ \vert h,q\ra = 0$  &\cr
\tablespace}\hrule}}
\hbox{\hskip 0.5cm Table 8:  ${\cal SW}(1,2)$ {\rm at } $c=12,Q=4$}}}
\mn
For the new algebras with $c\ge 3$ we obtained a very similar structure
for the representations. As for the $c=3d$ algebras the theories are not
rational and we also obtain a quantization of the $U(1)-$charge $q$.
\mn\sn
\vbox{
\line{\hfill\psfig{figure=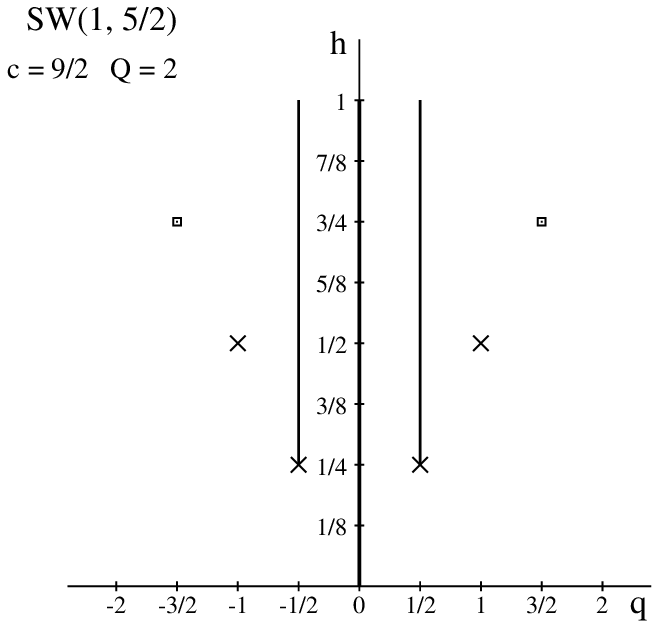}\quad
\quad \quad\psfig{figure=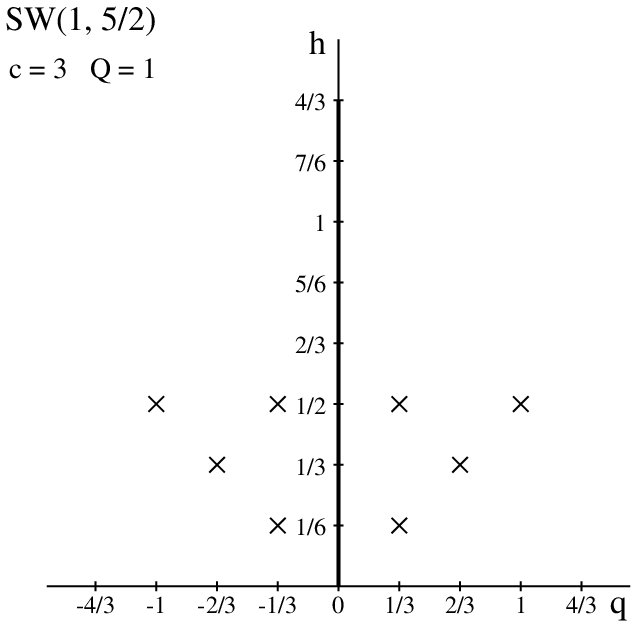}\hfill}
\vskip -4.5cm
\line{\hbox{$\scriptstyle
\vert {3\over 4},-{3\over 2}
\ra\sim\o{\psi}^-_0\vert {3\over 4},{3\over 2}\ra$}
\hskip 8.5cm
\hbox{}
\hfil}
\line{\hbox{$\scriptstyle
\vert h,-{1\over 2}\ra\sim{\psi}^-_0\vert h,
{1\over 2}\ra,\ h>{1\over 4}$}}
}
\vskip 4.5cm
\vfill\eject
\sn
The explicit data are given in the following two tables:
\bn
\centerline{\vbox{\hbox{\vbox{\offinterlineskip
\def\tablespace{height2pt&\omit&&\omit&\cr}
\def\tablerule{\tablespace\noalign{\hrule}\tablespace}
\hrule\halign{&\vrule#&\strut\hskip0.2cm\hfill#\hfill\hskip0.2cm\cr
\tablespace
& $(h,q)$ && \omit &\cr
\tablerule\tablerule
& $({1\over 2},\pm 1),\ (h,0)\quad {\rm for\ } 0\le h < \infty$ &&
$\la h,q\vert \o{\psi}_0^+ {\psi}_0^- \vert h,q\ra =
\la h,q\vert {\psi}_0^- \o{\psi}_0^+  \vert h,q\ra = 0$ &\cr
\tablespace
& \omit &&
$\la h,q\vert {\psi}_0^+ \o{\psi}_0^- \vert h,q\ra =
\la h,q\vert  \o{\psi}_0^- {\psi}_0^+\vert h,q\ra = 0$ &\cr
\tablerule
& $(h,+{1\over 2})\quad {\rm for\ } {1\over 4}\le h < \infty$ &&
$\la h,q\vert \o{\psi}_0^+ {\psi}_0^- \vert h,q\ra \sim (4h-1)^2,\
\la h,q\vert {\psi}_0^- \o{\psi}_0^+  \vert h,q\ra = 0$ &\cr
\tablespace
&\omit && $\la h,q\vert {\psi}_0^+ \o{\psi}_0^- \vert h,q\ra =
\la h,q\vert  \o{\psi}_0^- {\psi}_0^+\vert h,q\ra = 0$ &\cr
\tablerule
&  $(h,-{1\over 2})\quad {\rm for\ } {1\over 4}\le h < \infty$ &&
$\la h,q\vert {\psi}_0^- \o{\psi}_0^+  \vert h,q\ra\sim (4h-1)^2,\
\la h,q\vert \o{\psi}_0^+ {\psi}_0^- \vert h,q\ra = 0$ &\cr
\tablespace
& \omit && $\la h,q\vert {\psi}_0^+ \o{\psi}_0^- \vert h,q\ra =
\la h,q\vert  \o{\psi}_0^- {\psi}_0^+\vert h,q\ra = 0$ &\cr
\tablerule
& $({3\over 4}, +{3\over 2})$ && $\la h,q\vert
        {\psi}_0^+ \o{\psi}_0^- \vert h,q\ra \not= 0,\
        \la h,q\vert  \o{\psi}_0^- {\psi}_0^+\vert h,q\ra = 0$ &\cr
\tablespace
& \omit && $\la h,q\vert \o{\psi}_0^+ {\psi}_0^- \vert h,q\ra =
\la h,q\vert {\psi}_0^- \o{\psi}_0^+  \vert h,q\ra = 0$ &\cr
\tablerule
& $({3\over 4}, -{3\over 2})$ && $\la h,q\vert
        {\psi}_0^+ \o{\psi}_0^- \vert h,q\ra = 0,\
\la h,q\vert  \o{\psi}_0^- {\psi}_0^+\vert h,q\ra \not= 0$ &\cr
\tablespace
& \omit && $\la h,q\vert \o{\psi}_0^+ {\psi}_0^- \vert h,q\ra =
\la h,q\vert {\psi}_0^- \o{\psi}_0^+  \vert h,q\ra = 0$ &\cr
\tablespace}\hrule}}
\hbox{\hskip 0.5cm Table 9: ${\cal SW}(1,{5\over 2})$
     {\rm at } $c={9\over 2},Q=2$}}}
\mn\mn
\centerline{\vbox{\hbox{\vbox{\offinterlineskip
\def\tablespace{height2pt&\omit&&\omit&\cr}
\def\tablerule{\tablespace\noalign{\hrule}\tablespace}
\hrule\halign{&\vrule#&\strut\hskip0.2cm\hfill#\hfill\hskip0.2cm\cr
\tablespace
& $(h,q)$ && \omit &\cr
\tablerule\tablerule
& $(h,0)\quad {\rm for\ } 0\le h < \infty$ &&
$6h^3+5\la h,q\vert \o{\psi}_0^+{\psi}_0^-\vert h,q\ra=0,\
\la h,q\vert {\psi}_0^- \o{\psi}_0^+  \vert h,q\ra = 0 $  &\cr
\tablespace
& \omit && $\la h,q\vert {\psi}_0^+ \o{\psi}_0^- \vert h,q\ra =
\la h,q\vert  \o{\psi}_0^- {\psi}_0^+\vert h,q\ra = 0$ &\cr
\tablerule
& $({1\over 6},\pm{1\over 3}),({1\over 2},\pm{1\over 3}),
({1\over 3},\pm{2\over 3}),({1\over 2},\pm 1)$ &&
$\la h,q\vert \o{\psi}_0^+{\psi}_0^-\vert h,q\ra=
\la h,q\vert {\psi}_0^- \o{\psi}_0^+  \vert h,q\ra = 0 $ &\cr
\tablespace
& \omit &&  $\la h,q\vert {\psi}_0^+ \o{\psi}_0^- \vert h,q\ra =
\la h,q\vert  \o{\psi}_0^- {\psi}_0^+\vert h,q\ra = 0$  &\cr
\tablespace}\hrule}}
\hbox{\hskip 0.5cm Table 10: ${\cal SW}(1,{5\over 2})$
       {\rm at } $c=3,Q=1$ } }}
\bn\sn
In the following we will argue that all models with $c\ge 3$
can be understood from the existence of the spectral flow in
$N=2$ superconformal theories (see eq.\ (7)).
\sn
For $\eta\in \BZ$ the Neveu-Schwarz sector flows to itself.
Starting with the vacuum state $(h,q)=(0,0)$ and applying
successively the spectral flow with $\eta=\pm 1$ one obtains the
following tower of $N=2$ Virasoro primary states
\sn
$$  h_m={c\over 6}m^2-{(m-1)^2\over 2}, \quad
      q_m=\pm \left({mc\over 3}-m+1\right) \quad\quad m\in\BN.
\eqno{\rm (8)}$$
\sn
Note that this spectral flow does in general not map lowest weight
states onto lowest weight states but these subtleties can be easily
taken care of. For $c=3d$ it was shown in \q{\odz} that
the vacuum representation of the extension of the $N=2$ SVIR
by the local chiral primaries $(h,q)=(h_1=c/6,q_1=\pm c/3)$
is the direct sum of an infinite number of $N=2$ Virasoro
representations with lowest weights $(h_m, q_m)$.
This implies that the extension of the $N=2$ SVIR by the two fields
$(h,q)=(h_m,\pm \vert q_m \vert )$ also exists for $c=3d$ where
the generators can be written as normal ordered products of
the fundamental fields $(h_1,\pm\vert q_1\vert)$. Furthermore, in all
these theories the field ${\cal N}_s(\Phi{\cal L})$ vanishes.
\sn
Application of these arguments to $d=1$ implies, for instance, that all
${\cal SW}(1,\Dt)$ algebras with $\Dt=\BZ+1/2$ and $Q=1$ exist for $c=3$
being in agreement with our calculations.
Assuming that the construction of $\q{\odz}$ can be
generalized to arbitrary rational values of $c\ge 3$, yielding
generically a parafermionic fundamental algebra \q{\zam}, we
conjecture the ${\cal SW}(1,\Dt)$ algebra to exist for
$$  c={6\over m^2}\left( \Dt+{(m-1)^2\over 2}\right),\quad
 q=\pm {1\over m}(2\Dt-m+1)
\eqno{\rm (9)}$$
where $m\in\lbrace 1,\ldots, \lb \Dt+{1\over2} \rb\rbrace$.
\sn
We emphasize that the results of table 3 are in perfect agreement
with this conjecture.
\bn
\section{4.\ Conclusion}
\mn
Investigating the representation theory of $N=2$ $\w$-algebras with
two generators and vanishing self-coupling constant we found that
all known algebras can be arranged into four series. Only the algebras
existing for central charges in the unitary minimal series
of $N=2$ SVIR yield rational models which can be explained by
the non-diagonal modular invariants of $N=2$ SVIR. The $\w$-algebras
existing for $c\ge 3$ can be interpreted by the existence of the spectral
flow for $N=2$ supersymmetric CFTs. Every extension of the $N=2$ super
Virasoro algebra by the spectral flow operators implies a whole hierarchy
of $\w$-algebras existing for the same central charge. From our results
we expect that these algebras do not admit rational models. However, one
may expect a $U(1)-$charge quantization of the representations of these
algebras because this happened to be true in all cases considered so far.
In order to find new rational $N=2$ SCFTs one has to investigate
$N=2$ $\w$-algebras with more generators.
We rephrase that the only rational models known so far lie in the unitary
minimal series of $N=2$ SVIR. However, the question if unitarity is a
necessary condition for an $N=2$ supersymmetric CFT to be rational lies
beyond the scope of this letter but should
be investigated in the near future.
\bn
{\bf Acknowledgements:} We are very glad to thank
J.M.\ Figueroa-O'Farrill and W.\ Nahm for interesting
discussions and A.\ Honecker and R.\ Schimmrigk for
careful reading of the manuscript.
\bn\mn
\section{References}
\tolerance=10000
\mn
\bibitem{\bou} P.\ Bouwknegt, K.\ Schoutens,
       {\it $\w$-Symmetry in Conformal Field Theory},
        Phys.\ Rep.\ {\bf 223} (1993) p.\ 183

\bibitem{\ina} T.\ Inami, Y.\ Matsuo, I.\ Yamanaka,
{\it Extended Conformal Algebra with $N=2$ Supersymmetry},
Int.\ J.\ Mod.\ Phys.\ {\bf A5} (1990) p.\ 4441

\bibitem{\lup} H.\ Lu, C.N.\ Pope, L.J.\ Romans, X.\ Shen, X.J.\ Wang,
{\it Polyakov Construction of the $N=2$ Super-${\cal W}_3$ Algebra},
Phys.\ Lett.\ {\bf B264} (1991) p.\ 91

\bibitem{\rom} L.J.\ Romans, {\it The $N=2$ Super-${\cal W}_3$ Algebra},
Nucl.\ Phys.\ {\bf B369} (1992) p.\ 403

\bibitem{\nem} D.\ Nemeschansky, S.\ Yankielowicz,
{\it $N=2$ $\w$-Algebras, Kazama-Suzuki Models and Drinfeld Sokolov
Reduction}, preprint USC-91/005

\bibitem{\ito} K.\ Ito, {\it Quantum Hamiltonian Reduction
and $N=2$ Coset Models}, Phys.\ Lett.\ {\bf B259} (1991) p.\ 73

\bibitem{\blm} R.\ Blumenhagen, {\it $N=2$ Supersymmetric $\w$-Algebras},
Nucl.\ Phys.\ {\bf B405} (1993) p.\ 744

\bibitem{\ban} T.\ Banks, L.J.\ Dixon, D.\ Friedan, E.\ Martinec,
{\it Phenomenology and Conformal Field Theory Or
     Can String Theory predict the Weak Mixing Angle?},
Nucl.\ Phys.\ {\bf B299} (1988) p.\ 613

\bibitem{\gep} D. Gepner, {\it Space-time Supersymmetry in
Compactified String Theory and Superconformal Models},
Nucl.\ Phys.\ {\bf B296} (1988) p.\ 757

\bibitem{\kaz} Y.\ Kazama, H.\ Suzuki, {\it New $N=2$
Superconformal Field Theories and Superstring Compactification},
Nucl.\ Phys.\ {\bf B321} (1989) p.\ 232

\bibitem{\ber} M.\ Bershadsky, W.\ Lerche, D.\ Nemeschansky, N.P.\ Warner,
{\it Extended $N=2$ Superconformal Structure of Gravity and
$\w$-Gravity Coupled to Matter}, Nucl.\ Phys.\ {\bf B401} (1993) p.\ 304

\bibitem{\var} R.\ Varnhagen, {\it Characters and Representations of
New Fermionic $\w$-Algebras}, Phys.\ Lett.\ {\bf B275} (1992) p.\ 87

\bibitem{\ehe} W.\ Eholzer, M.\ Flohr, A.\ Honecker, R.\ H\"ubel,
W.\ Nahm, R.\ Varnhagen, {\it Representations of $\w$-Algebras
with Two Generators and New Rational Models},
Nucl.\ Phys.\ {\bf B383} (1992) p. 249

\bibitem{\rav} F.\ Ravanini, S.-K.\ Yang, {\it Modular Invariance
in $N=2$ Superconformal Field Theories},
Phys.\ Lett.\ {\bf B195} (1987) p.\ 202

\bibitem{\qiu} Z.\ Qiu, {\it Modular Invariant Partition
Functions for $N=2$ Superconformal Field Theories},
Phys.\ Lett.\ {\bf B198} (1987) p.\ 497

\bibitem{\odz} S.\ Odake, {\it $c=3d$ Conformal Algebra with
Extended Supersymmetry}, Mod.\ Phys.\ Lett.\ {\bf A5} (1990) p.\ 561

\bibitem{\egu} T.\ Eguchi, H.\ Ooguri, A.\ Taormina, S.-K.\ Yang,
               {\it Superconformal Algebras and String
                Compactification on Manifolds with $SU(n)$ Holonomy},
                Nucl.\ Phys.\ {\bf B315} (1989) p.\ 193

\bibitem{\wes} P.\ West, {\it 'Introduction to Supersymmetry
and Supergravity'}, second edition 1990, World Scientific, Singapore

\bibitem{\mus} G.\ Mussardo, G.\ Sotkov, M.\ Stanishkov,
{\it $N=2$ Superconformal Minimal Models},
Int.\ J.\ Mod.\ Phys.\ {\bf A4} (1989) p.\ 1135

\bibitem{\hon} A.\ Honecker, {\it A Note on the Algebraic
Evaluation of Correlators in Local Chiral Conformal Field Theory},
preprint BONN-HE-92-25, hep-th/9209029

\bibitem{\gin} P.\ Ginsparg, {\it Curiosities at $c=1$},
               Nucl.\ Phys.\ {\bf B295} (1988) p.\ 153

\bibitem{\kir} E.B.\ Kiritsis, {\it Proof of the Completeness of the
               Classification of Rational Conformal Theories with $c=1$},
               Phys.\ Lett.\ {\bf B217} (1989) p.\ 427

\bibitem{\flo} M.\ Flohr, {\it $\w$-Algebras, New Rational Models
and Completeness of the $c=1$ Classification},
Commun.\ Math.\ Phys.\ {\bf 157} (1993) p.\ 179

\bibitem{\zam} V.A.~Fateev, A.B.~Zamolodchikov,
{\it Nonlocal (Parafermion) Currents in Two-Dimensional Conformal
Quantum Field Theory and Self-Dual Critical Points
in $\BZ_N$-Symmetric Statistical Systems},
Sov.\ Phys.\ JETP {\bf 62} (1985) p.\ 215

\vfill
\end